\documentclass[aps,twocolumn, floats,preprintnumbers,showpacs,nofootinbib,prd]{revtex4-1}
\pdfoutput=1
\usepackage{hyperref}
\usepackage{graphicx}
\usepackage{url}
\usepackage{amsmath}
\usepackage{amsfonts}
\usepackage{amssymb}
\def\beq{\begin{equation}}
\def\eeq{\end{equation}}
\def\beqa{\begin{eqnarray}}
\def\eeqa{\end{eqnarray}}

\def\ltap{\ \raise.3ex\hbox{$<$\kern-.75em\lower1ex\hbox{$\sim$}}\ }
\def\gtap{\ \raise.3ex\hbox{$>$\kern-.75em\lower1ex\hbox{$\sim$}}\ }

\begin{document}

\title{Natural Milli-Charged Inflation
}

\author{Yang Bai and Ben A. Stefanek
\\
\vspace{2mm}
\normalsize\emph{Department of Physics, University of Wisconsin, Madison, WI 53706, USA} 
}


\begin{abstract}
We construct a natural inflation model with the inflaton as a linear combination of the fifth components of Abelian gauge fields in a five-dimensional theory. A seesaw mechanism is introduced to provide a natural milli-charge for matter fields under one combination of the gauge symmetries. As a result, the effective decay constant of the inflaton field can be above the Planck scale with all scales in the model below the Planck scale. Our model predicts a tensor-to-scalar ratio $r$ between 0.033 and 0.125 for sixty $e$-folds and a reheating temperature of a few $10^{11}$~GeV.
\end{abstract}
\maketitle

\noindent
{\it{\textbf{Introduction.}}}
Natural inflation provides an elegant way to incorporate the general inflation idea to solve many problems in the big-bang theory~\cite{Freese:1990rb}. Treating the inflaton as a Pseudo Nambu-Goldstone Boson (PNGB), the flatness of the inflaton potential can be protected by a shift symmetry from a spontaneously broken symmetry. The current experimental results from WMAP~\cite{Hinshaw:2012aka} and Planck~\cite{Ade:2013uln} have started to constrain the natural inflation model parameter space. The decay constant of the inflaton PNGB is required to be trans-planckian, which makes the effective field theory description unreliable. An additional mechanism is therefore required to have all mass scales to be below the Planck scale. 

One of the mechanisms is the so-called aligned axion mechanism in Ref.~\cite{KNP2004}, where two axion fields have two cosine function potential terms and there are two decay constants for each axion. If the two ratios of the two pairs of decay constants approximately equal to each other, an effective decay constant above the Planck scale can be achieved even though all the original decay constants are sub-Planckian. The inflaton field can be identified as the linear combination of the two axions with a large effective decay constant. In the field space, the inflaton can travel in a helical trajectory with trans-Planckian change of field values during inflation~\cite{TW2014, CKY2014, KKN2014}. This mechanism seems to solve the problem. However, without a natural realization of the relation of the few decay constants, this aligned axion model still requires tuning in the parameter space. In this paper, we provide a model to naturally realize the aligned feature. 

More specifically, we will treat the PNGB's as the fifth components of gauge fields propagating in a five dimensional spacetime, following a similar setup as in Ref.~\cite{ArkaniHamed:2003wu}. The shift-symmetry of the PNGB in the four dimensional theory can be matched to a gauge symmetry in 5D. To have more than one PNGB's, we will simply introduce a product of $U(1)$'s in the 5D theory. At tree level, there is no potential for all PNGB's. At the one-loop level, the matter fields propagating in the bulk can generate a non-local potential for the gauge-invariant Wilson loop~\cite{YH1983, HIL1998, Antoniadis:2001cv, GIQ2002}. Multiple cosine function terms then appear in the effective potential with various decay constants relating to different matter charges. In this specific setup, the effective large decay constant can be translated into a milli-charge of one matter field under one linear combination of gauge symmetries. 

As pointed out in Ref.~\cite{BH1986, SSY2013}, milli-charges can exist for matter fields under a spontaneously broken gauge theory from kinetic mixings or if there are more than one unbroken $U(1)$'s. Naively speaking, one could try to use a small value of the kinetic mixing term to realize a large effective decay constant. However, the field redefinition procedure can not change the rank of a matrix. In other words, the initial charge matrix has to have a small determinant to eventually provide a milli-charge for one combination of gauge groups. Noticing this fact, we introduce a seesaw structure for the charge matrix and  obtain a tiny determinant with modestly hierarchical charge values. The seesaw structure could come from some ultra-violet (UV) models. We will not explore its detailed UV realization  in this paper.

\noindent
{\it{\textbf{Effective Potential of the Fifth Components.}}}
\vspace*{1mm}
Following a similar setup as the extra-natural inflation in Ref.~\cite{ArkaniHamed:2003wu}, we consider gauge fields as well as their matter fields propagating in 5D spacetime, with a circular fifth dimension in the range of $(0, 2\pi R)$. We keep our setup as general as possible by studying a product of $N$ $U(1)$'s: $U(1)_1 \times U(1)_2 \times \cdots U(1)_N$. The number of matter fields charged under these $U(1)$'s is required to be greater than or equal to $N$ such that the fifth component of all the gauge fields can become massive at the one-loop level. For simplicity, we choose the number of matter fields to also be $N$. The charges of the matter fields can be described by an $N\times N$ matrix, which we denote as $Q_{a i}$ with $a, i= 1, \cdots\,N$ for the matter index ``$a$" and gauge index ``$i$". 

Since the effective five dimensional theory has the fifth dimension compactified on a circle, we have the relation between the 5D Planck scale $M_5$ and the 4d reduced Planck scale $M_{\rm pl} = 1/ \sqrt{8\pi G_N} = 2.4 \times 10^{18}$~GeV as $M_{\rm pl}^2 = 2\pi R M_5^3$. To have a controllable effective field theory, the compactification scale is required to be parametrically smaller than the 5D cutoff $1/R \ll M_5$. Requiring at least three (two) KK-modes below the 5D Planck scale, we need to have $1/R < 0.13\,(0.2)~M_{\rm pl}$. 

For an individual matter field, its KK-modes can generate the effective Coleman-Weinberg potential for the Wilson line of $\oint dx_5  \sum_{i} Q_{a i} g_i A_{i,5}$, with $g_i$ as gauge couplings. Adding all matter contributions together, we have the effective potential as~\cite{Antoniadis:2001cv}
\beqa
&& \hspace{-0.1cm} V_{\rm eff} (A_{1,5}, A_{2,5}, \cdots, A_{N,5}) =    \\
&& \hspace{0.1cm} V_0  - \frac{3}{64\pi^6\,R^4} \sum_{a =1}^N (-1)^{F_a}  \sum_{i = 1}^N \sum_{n=1}^\infty \frac{\cos{(2\pi R n Q_{a i} g_i A_{i,5}) }}{n^5}  \, ,\nonumber
\eeqa
where $F_a=0(1)$ for massless bosonic (fermionic) matter field (see Ref.~\cite{Feng:2003mk} for the massive matter case). The constant $V_0$ is introduced such that the potential is zero at its minimum. Since the potential is dominated by the lightest KK-mode with $n=1$, we approximately end up with $N$ cosine functions with different frequencies.  The decay constant matrix is the inverse of the frequency matrix and is
\beqa
f_{{\rm decay}, a i} = \frac{1}{2\pi R}\, \left[ Q_{a i } g_i \delta_{ij} \right]^{-1} \,.
\eeqa
So to obtain one decay constant much above the 4d Planck scale, we need to have a small eigenvalue for the matrix $Q_{a i } g_i \delta_{ij}$. One way to achieve this is to have tiny gauge couplings for all gauge bosons~\cite{ArkaniHamed:2003wu}, which will make all eigenvalues to be small. Another way is to have a tiny determinant for the charge matrix $Q_{a i}$, which will only make some of the eigenvalues tiny and will be the concentration of this paper.  

As from the familiar story for the neutrino masses from the seesaw mechanism, the lightest eigenvalue can be more suppressed compared to the generic small parameters in the matrix. By choosing a seesaw structure for the matter charge matrix, we can have the lightest eigenvalue of $f^{-1}_{\rm decay}$ generically smaller than all other eigenvalues. For the inflation purpose and in the basis of $A^\prime_{i, 5}$ with a diagonal decay constant matrix, we will have the inflation-relevant decay constant denoted as $A^\prime_{1, 5}$, under which at least one matter field can a tiny or milli- charge. Assuming that the effective potential has already evolved to reach the minimum points in other orthogonal directions, we can identify the inflaton as $A^\prime_{1, 5}$ and have the effective decay constant as
\beqa
f_{\rm inflaton} = \mbox{Max}[ {\rm Eigenvalues}( f_{\rm decay} ) ] \,.
\eeqa
It is fairly simple to see that the effective inflaton decay constant is enhanced by $1/\mbox{det}(Q)$. 

\noindent
{\it{\textbf{Two $U(1)$'s and Three $U(1)$'s.}}}
\vspace*{1mm}
Using two $U(1)$'s as a concrete example, we choose the charge matrix of two matter fields to have a seesaw formula 
\beqa
Q = 
\renewcommand{\arraystretch}{1.3}
\left(
\begin{matrix}
 0   &   -q_1    \\
 q_1  &  q_2   
\end{matrix}
\right) \,,
\eeqa
with $q_1 \ll q_2$.~\footnote{We don't need to have a really tiny ratio of $q_1/q_2$, but we do require $q_1$ to be parametrically smaller than $q_2$.} Diagonalizing this matrix, we have the large decay constant as
\beqa
f_{\rm inf} = \frac{q_2}{2\pi R\,g\,q_1^2} \equiv  \frac{1}{2 \pi R\, q_{\rm eff} } \,,
\eeqa
for $g_1=g_2\equiv g$. A large tensor-to-scalar ratio $r$ requires a decay constant above $\sim 10\,M_{\rm pl}$ for a broad class of natural inflation models~\cite{Freese:2014nla}. For instance, choosing $R^{-1} = 0.1\,M_{\rm pl}$, $g=0.3$, $q_2=1$ and $q_1=1/6$, we have $q_{\rm eff} = 0.0083$ and $f_{\rm inf} \approx 1.9~M_{\rm pl}$, so we may still have some difficulty in obtaining a large enough $r$. 

If there are three $U(1)$'s, the seesaw formula can have the form
\beqa
Q = 
\renewcommand{\arraystretch}{1.3}
\left(
\begin{matrix}
 0   &   q_1 & 0   \\
 q_1  &  0  & q_3  \\
 0    &    q_3  & q_2 \\  
\end{matrix}
\right) \,,
\eeqa
with $q_1, q_2 \ll q_3$. The largest decay constant has
\beqa
f_{\rm inf} = \frac{q_3^2}{2\pi R\,g\,q_1^2\,q_2}  \equiv  \frac{1}{2 \pi R\, q_{\rm eff} }  \,, 
\label{eq:Finf}
\eeqa
again for $g_1=g_2=g_3=g$. Choosing a benchmark parameter point, $R^{-1} = 0.1\,M_{\rm pl}$, $g=0.3$, $q_3=1$ and $q_1=q_2=1/6$, we have $q_{\rm eff} = 0.0014$ and $f_{\rm inf} \approx 11.4~M_{\rm pl}$, which may be sufficient to obtain a large $r$.

\noindent
{\it{\textbf{Inflation Parameters.}}}
\vspace*{1mm}
The general feature of the inflation potential is similar to a single cosine function potential from a natural inflation model~\cite{Freese:1990rb}. Neglecting the other PNGB's, we have the potential in terms of the lighter inflaton given by
\beqa
V_{\rm eff} =  \frac{3}{64\pi^6\,R^4}   \sum_{n=1}^\infty \frac{1}{n^5}  \, \left[ 1  - \cos \left({ \frac{n A_{1, 5}^\prime}{f_{\rm inf} }}\right) \right]   \,,
\eeqa
for bosonic degrees of freedom of matter fields. The slow-roll parameters are calculated to be~\cite{Baumann:2009ds}
\beqa
\epsilon &=& \frac{M_{\rm pl}^2}{2} \left( \frac{V^\prime_{\rm eff}}{V_{\rm eff} } \right)^2  \approx \frac{M_{\rm pl}^2}{2 f^2_{\rm inf} } \cot^2\left( \frac{A_{1,5}^\prime}{2 f_{\rm inf} } \right) \sim \frac{M_{\rm pl}^2}{f^2_{\rm inf} }  \,, \nonumber \\
\eta &=& M_{\rm pl}^2 \, \frac{V^{\prime\prime}_{\rm eff}}{V_{\rm eff} }   \approx \frac{M_{\rm pl}^2}{2 f^2_{\rm inf} } \cos\left( \frac{A_{1,5}^\prime}{f_{\rm inf} } \right) \csc^2\left( \frac{A_{1,5}^\prime}{2 f_{\rm inf} } \right)  \sim \frac{M_{\rm pl}^2}{f^2_{\rm inf} }  \,, \nonumber \\
A_s &=& \frac{V}{24\pi^2 M^4_{\rm pl} \epsilon} \sim  \frac{f_{\rm inf}^2 }{512 \pi^{10} M_{\rm pl}^6 R^4}  \,, \nonumber 
\eeqa
and the spectral index $n_s = 1 - 6 \epsilon + 2 \eta$ and the tensor-to-scalar ratio $r=16 \epsilon$. 

To obtain numerical values of $A_s$, $n_s$ and $r$, one needs to choose the initial value of $A^\prime_{1, 5}$. The initial value, $A^{\prime, {\rm init}}_{1, 5}$, is determined by the number of $e$-folds 
\beqa
N_{\rm e}  \approx \frac{1}{M_{\rm pl} } \int^{A^{\prime, {\rm init}  }_{1, 5} }_{A^{\prime, {\rm end} }_{1, 5}} \frac{d A^\prime_{1, 5} }{ \sqrt{2 \epsilon} } \,,
\eeqa
with the end of inflation at $\epsilon(A^{\prime, {\rm end} }_{1, 5}) \approx 1$. For the benchmark point with $f_{\rm inf} \approx 11.4\,M_{\rm pl}$, we have $A^{\prime, {\rm end} }_{1, 5} \approx 1.4\,M_{\rm pl}$ and $A^{\prime, {\rm init} }_{1, 5} \approx 15.0(13.8)\,M_{\rm pl}$ for $60 (50)$ $e$-folds. The scalar amplitude is $A_s = 2.34 (1.62) \times 10^{-9}$ and the spectral indexes are $n_s = 0.966 (0.960)$ and $r = 0.10 (0.13)$ for $60 (50)$ $e$-folds.

\begin{figure}[th!]
\begin{center}
\hspace*{-0.0cm}
\includegraphics[width=0.48\textwidth]{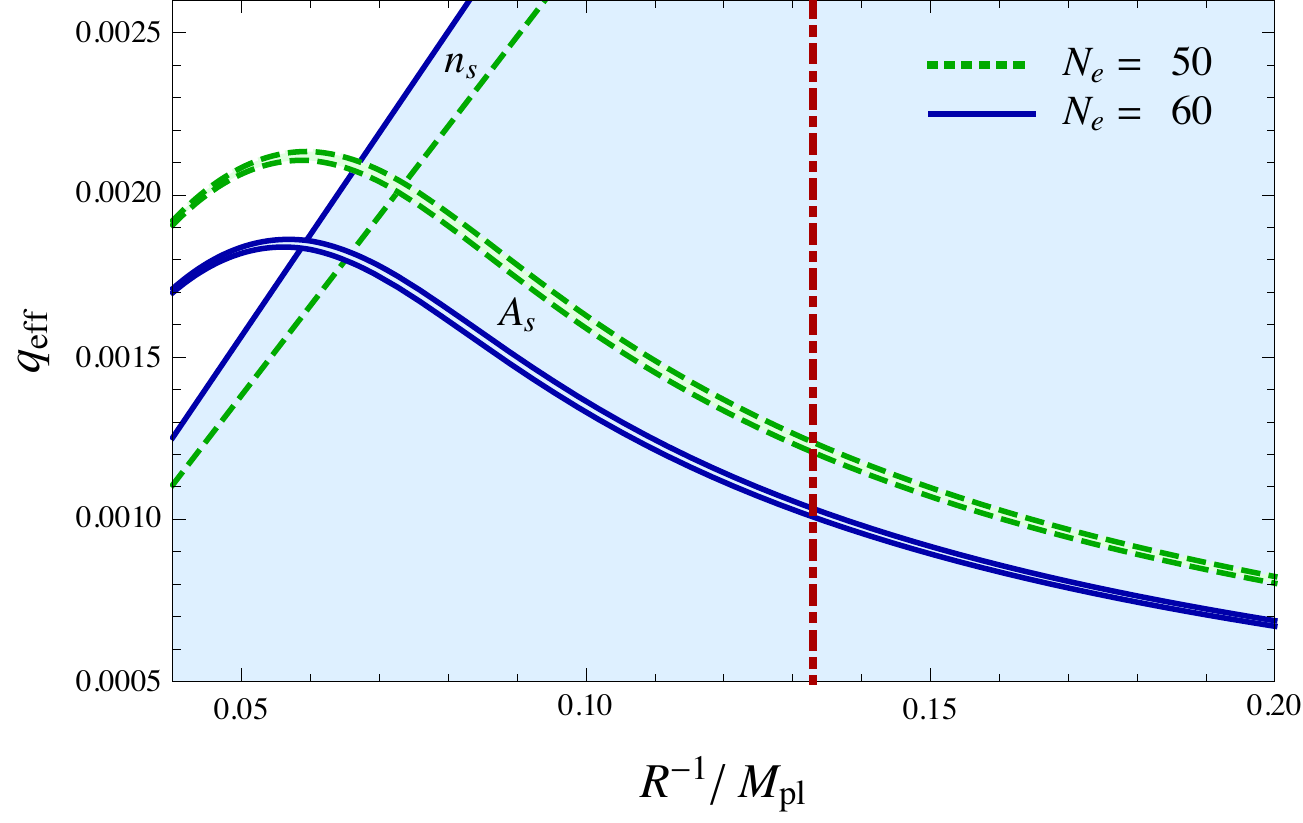}
\caption{The allowed model parameter space in the effective matter charge $q_{\rm eff}$ and the compactification scale $R^{-1}$ compatible with the experimentally measured slow-roll parameters $A_s$ and $n_s$. The red and dotdashed line is the bound for having three KK-modes below the 5D cutoff $M_5$. 
}
\label{fig:charge-r}
\end{center}
\end{figure}

\begin{figure}[th!]
\begin{center}
\hspace*{-0.0cm}
\includegraphics[width=0.48\textwidth]{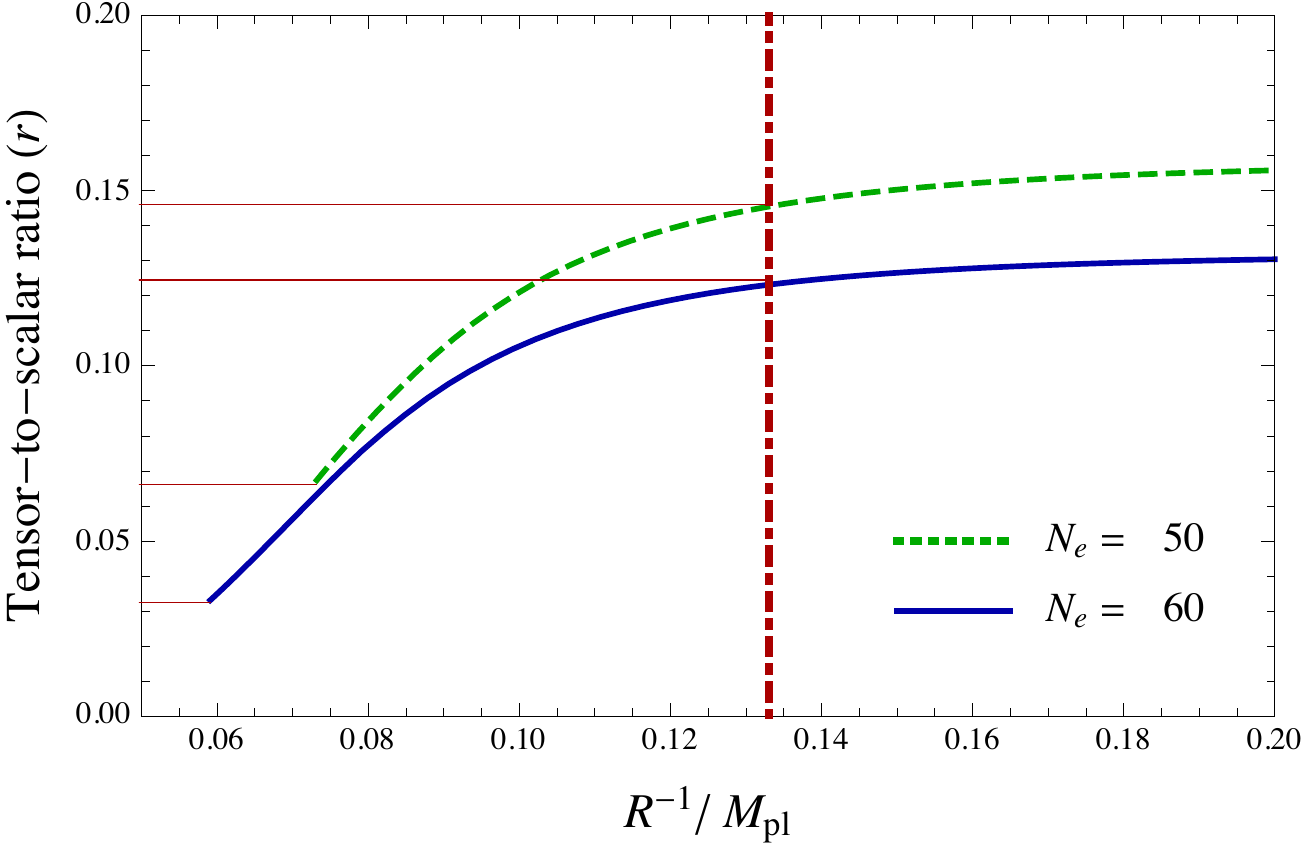}
\caption{The tensor-to-scalar ratio $r$ as a function of $R^{-1}$ after satisfying the measured values of $A_s$ and $n_s$. The upper values of $r$ for $60 (50)$ $e$-folds are shown in red lines and the lower values of $r$ are at start points of the lines.}
\label{fig:r-Rinv}
\end{center}
\end{figure}

In our model, we have only two parameters: $R$ and $q_{\rm eff}$, relevant for inflation. The measured values of $A_s=2.196^{+0.053}_{-0.059}\times 10^{-9}$ and $n_s = 0.9603\pm 0.0073$ from Planck collaboration in Ref.~\cite{Ade:2013uln} can determine all our model parameters. In Fig.~\ref{fig:charge-r}, we show the allowed parameter space after satisfying the 1$\sigma$ band of measured values of $A_s$ and $n_s$. The scalar amplitude $A_s$ provides a non-trivial dependence of $q_{\rm eff}$ on $R^{-1}$, while the primordial tilt $n_s$ only depends on $f_{\rm inf}$ and thus is linear in the $q_{\rm eff}$ and $R^{-1}$ space. For a given $q_{\rm eff}$, $n_s$ provides a lower limit of $R^{-1}/M_{\rm pl} > 0.0591 (0.0732)$ for $60 (50)$ $e$-folds, but not an upper bound. This is because for a large value of $R^{-1}$ or $f_{\rm inf}$, the natural inflation model reaches the $m^2 \phi^2$ potential and has $n_s$ inside the $1\sigma$ band of $n_s$. To trust the effective 5D field description, we further require $R^{-1}/M_{\rm pl} < 0.133$ to have at least three KK-modes below the 5D cutoff $M_5$. After satisfying the constraints of $A_s$ and $n_s$, we show the model predictions for the tensor-to-scalar ratio $r$ in Fig.~\ref{fig:r-Rinv}. For $60 (50)$ $e$-folds, we have the preferred values of $r$ to be $0.033< r < 0.125$ $(0.066 < r < 0.146)$. 

\begin{figure}[th!]
\begin{center}
\hspace*{-0.0cm}
\includegraphics[width=0.48\textwidth]{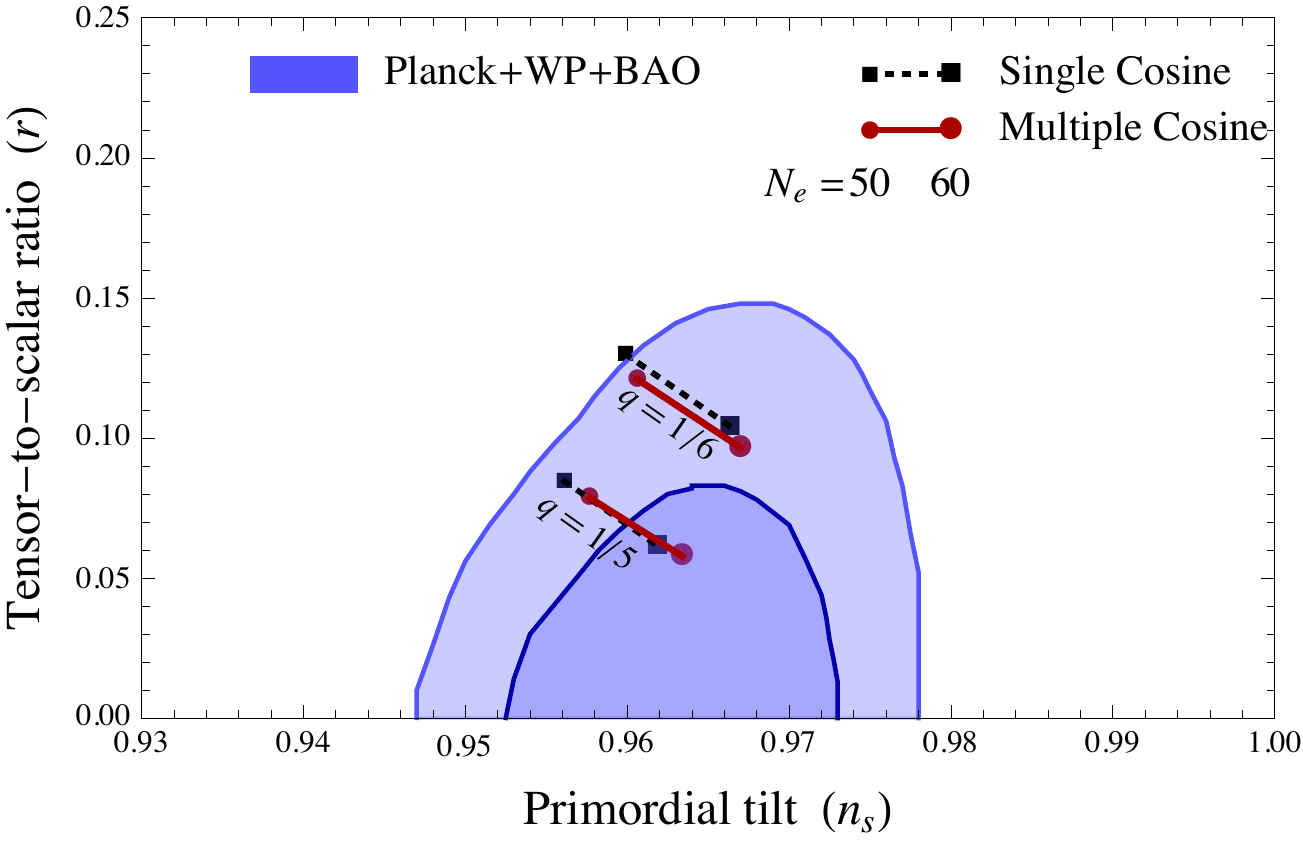}
\caption{Spectral indexes, $n_s$ and $r$, are shown in the red and circular points for the natural milli-charged inflation model. The model parameters are chosen to have $R^{-1}=0.1\,M_{\rm pl}$, $g=0.3$, $q_3 = 1$ and $q_1=q_2=q$. As a comparison, we also show the original natural inflation model with only one cosine function in the black and square points. The blue shaded regions are constraints from Planck+WMAP+BAO with $68\%$ and $95\%$ CL~\cite{Ade:2013uln}.}
\label{fig:ns-r}
\end{center}
\end{figure}

To convert the effective charge $q_{\rm eff}$ to the basic parameters in our model, we show the model points in the $n_s - r$ plane for different choices of charges in Fig.~\ref{fig:ns-r} by keeping $R^{-1}=0.1\,M_{\rm pl}$ and $g=0.3$. Comparing the results for $N_e=50$ and $N_e=60$, one can see that a large number of $e$-folds yields a smaller value of $r$ but a larger value of $n_s$. Our extra-dimensional motivated model has different trajectories of model points from the simplest natural inflation model. For a fixed lowest frequency, our model as well as the extra-natural inflation model prefer slightly smaller values of $r$ and larger values of $n_s$. For a comparison to experimental data, we show the current experimental constraints from the Planck collaboration by including Planck CMB temperature data, WMAP large-scale polarization data and baryon acoustic oscillations (BAO) data~\cite{Ade:2013uln}.

The recent $B$-mode measurement from the BICEP2 experiment has been interpreted as the detection of primordial gravity waves with the tensor-to-scalar ratio $r = 0.20^{+0.07}_{-0.05}$~\cite{BICEP2}, although additional dust polarization could change this conclusion~\cite{Ade:2014gna, MS2014}. Our model will be falsified if the constraints on $r$ can be below around 0.05 from future experimental measurements. 

\noindent
{\it{\textbf{Reheating Temperature.}}}
\vspace*{1mm}
At the end of the inflation, the inflaton can decay into lighter matter fields and reheat the Universe. In our model, we have  matter fields charged under the gauge symmetry $U(1)^\prime_1$ with a milli-charge $q_{\rm eff}$, so the large decay constant of the inflaton field can be kept. For a fermonic matter field $\Psi$ in 5D, we have the coupling of the inflaton field $A^\prime_{1, 5}$ to two fermions as $i\,q_{\rm eff} \, A^\prime_{1, 5} \, \overline{\psi} \gamma_5 \psi$ with $\psi$ as the zeroth mode of $\Psi$.~\footnote{In the minimal model, the inflaton can not decay into two scalar zeroth modes, as can be seen directly from the 5D coupling ${\cal A}_5 (\partial_5 \Phi) \Phi^\dagger$ from the replacement of $\partial_5 \rightarrow i\,n/R$.} The inflaton decay width is 
\beqa
\Gamma( A^\prime_{1, 5} \rightarrow \overline{\psi} \psi ) = \frac{q_{\rm eff}^2 }{8\pi} \,M_{A^\prime_{1, 5} } \,,
\eeqa
where the inflaton mass around the minimum of potential is calculated to be
\beqa
M_{A^\prime_{1, 5} }^2 = \frac{3}{64\pi^6 R^4} \sum_{n=1}^{\infty} \frac{1}{n^5} \left(\frac{n}{f_{\rm inf}} \right)^2 = 
\frac{3\,\zeta(3) }{64\pi^6 R^4 f_{\rm inf}^2}   \,,
\eeqa
with $\zeta(3)\approx 1.2$. 

The reheating temperature, $T_{\rm rh}$, can be estimated when the Hubble parameter $H$ reaches the decay width of the inflaton, at which point the universe is in thermal equilibrium. Approximately, $T_{\rm rh}$ is~\cite{Bassett:2005xm}
\beqa
\hspace{-0.2cm} T_{\rm rh} \simeq 0.2 \left( \frac{100}{g_*} \right)^{1/4} \sqrt{  \sqrt{8\pi} \,\Gamma M_{\rm pl} } \approx 0.02 \, \sqrt{\frac{M_{\rm pl}\, q_{\rm eff}^3}{R} }\,, 
\eeqa
for the radiation degrees of freedom $g_* = {\cal O}(100)$. Using the constraints from $A_s$ and $n_s$ in Fig.~\ref{fig:charge-r}, we have $T_{\rm rh} \simeq 5.6 - 9.3 (7.3 - 11.5) \times 10^{11}$~GeV for 60(50) $e$-folds. This value of reheating temperature can affect the parameter space of the leptogenesis and baryogenesis models. 

\vspace*{0.3cm}
\noindent
{\it{\textbf{Discussion and Conclusions.}}}
As pointed out in Ref.~\cite{BDFG2003}, the string theory seems unlikely to realize a large effective decay constant for the natural inflation model. A more general argument has been provided from the weak gravity conjecture~\cite{AHMNV2006} motivated by arguments involving holography, the absence of Planck scale remnants and the incompatibility of global symmetries with quantum theories of gravity. It states that in an effective 4d theory with gravity and a $U(1)$ gauge field with a small gauge coupling or only mill-charged matters, there exist a ``hidden cutoff", $\Lambda_{\rm hidden} = q_{\rm eff} M_{\rm pl} \sim 10^{-3}\,M_{\rm pl}$, in the notation of our model. As can be seen from Fig.~\ref{fig:charge-r}, the current experimental data prefer to have the compactification scale $1/R \gtrsim 0.06\,M_{\rm pl}$ and above $\Lambda_{\rm hidden}$, so the weak gravity conjecture is violated from the 4d point of view. For the 5D theory, no tiny charges have been introduced in the original basis of $U(1)$'s. However, unless there are some other properties to distinguish those $U(1)$'s, one still has milli-charge from the beginning in the rotated basis. One may argue that gauge boson masses at lower energy scales can distinguish those $U(1)$'s without disturbing the inflation part. Unfortunately, the hierarchic scale separations between gauge boson masses and the compactification scale regenerate fine-tuning problems for the Higgs fields charged under $U(1)$'s, similar to the ``hierarchy problem" of the electroweak sector in the Standard Model. 

In summary, we have constructed and analyzed a natural inflation model with multiple PNGB's as fifth components of 5D Abelian gauge bosons. A seesaw structure of the charge matrix is introduced to obtain a milli-charge for one linear combination of gauge fields. A trans-Planckian effective decay constant has been achieved with all scales kept sub-Planckian. The tensor-to-scalar ratio is preferred to be in the range of $0.033< r < 0.125$ $(0.066 < r < 0.146)$ for 60(50) $e$-folds, while the reheating temperature is a few $10^{11}$~GeV.

\vspace{3mm}
{\it{\textbf{Acknowledgements.}}}
We thank Haipeng An, Hsin-Chia Cheng, Aki Hashimoto, Josh Ruderman, Henry Tye and Xin-min Zhang  for useful discussion.  This work is supported by the U. S. Department of Energy under the contract DE-FG-02-95ER40896. YB thanks the Center for Future High Energy Physics, where this work is finished. 


\providecommand{\href}[2]{#2}\begingroup\raggedright\endgroup

\end{document}